\documentclass{rspublicx}
\usepackage{dsfont}
\usepackage[dvips]{graphicx}
\usepackage{amssymb,amsfonts}
\usepackage{amsmath,exscale,amsthm}

\input cyracc.def 
  \font\tencyr=wncyr10
  \def\cyr{\tencyr\cyracc}

\begin{document}

\newtheorem{lem}{Lemma}
\newtheorem{prop}{Proposition}
\newcommand{\half}{\mbox{$\textstyle \frac{1}{2}$}}
\newcommand{\quat}{\mbox{$\textstyle \frac{1}{4}$}}

\title[Informed Traders]{Informed Traders}

\author[Brody, Davis, Friedman \& Hughston]{Dorje~C.~Brody${}^1$,
Mark~H.~A.~Davis${}^{1}$, Robyn~L.~Friedman${}^{1,2}$ \\ and
Lane~P.~Hughston${}^1$}

\affiliation{${}^1$Department of Mathematics, Imperial College
London, London SW7 2BZ, UK \\ ${}^2$Royal Bank of Scotland, 135
Bishopsgate, London EC2M 3UR, UK }

\date{\today}
\maketitle

\begin{abstract}{statistical arbitrage; asymmetric information;
information-based asset pricing; hedge funds; insider trading;
mutual information} An asymmetric information model is introduced
for the situation in which there is a small agent who is more
susceptible to the flow of information in the market than the
general market participant, and who tries to implement strategies
based on the additional information. In this model market
participants have access to a stream of noisy information concerning
the future return of an asset, whereas the informed trader has
access to a further information source which is obscured by an
additional noise that may be correlated with the market noise. The
informed trader uses the extraneous information source to seek
statistical arbitrage opportunities, while at the same time
accommodating the additional risk. The amount of information
available to the general market participant concerning the asset
return is measured by the mutual information of the asset price and
the associated cash flow. The worth of the additional information
source is then measured in terms of the difference of mutual
information between the general market participant and the informed
trader. This difference is shown to be nonnegative when the
signal-to-noise ratio of the information flow is known in advance.
Explicit trading strategies leading to statistical arbitrage
opportunities, taking advantage of the additional information, are
constructed, illustrating how excess information can be translated
into profit.
\end{abstract}
\maketitle

\section{Introduction}

There are many different approaches to the modelling of so-called
insider trading strategies. Starting with the work of Kyle (1985)
and Back (1992), a number of investigations have been carried out
(to name a few, Amendinger \textit{et al}. 1998, Seyhun 1988,
F\"ollmer \textit{et al}. 1999, Back \textit{et al}. 2000, Le\'on
\textit{et al}. 2003, Corcuera \textit{et al}. 2004, Biagini \&
{\O}ksendal 2005, 2006, Ankirchner \textit{et al}. 2006, Campi \&
\c{C}etin 2007). It is sometimes assumed in the literature that the
`insider' has direct access to the values of future asset prices.
While such a scenario may indeed occasionally prevail, the more
common situation is that informed agents do not have advance access
to the exact values of future asset prices. How, then, do agents
having high information susceptibility---when compared to the
average market participant---utilise their strengths in reality? An
increasingly popular strategy adopted by some large hedge funds is
to make use of publicly available information in addition to
high-frequency price data. What gives these funds an edge is their
significant computational power for data and text mining, thus
allowing them to extract useful information from publicly available
sources \textit{faster} than their competitors. Against this
background it is natural to ask how much added information an extra
source provides, how does it affect trading strategies, and more
generally in what way can such information-based strategies be
modelled mathematically.

The purpose of the present paper is: (i) to introduce a
phenomenological approach to model the agent susceptive to
information, (ii) to quantify the amount of added information, and
(iii) to derive trading strategies that lead to statistical
arbitrage opportunities for such an agent. Our analysis is carried
out within the information-based asset pricing framework of Brody,
Hughston and Macrina (Macrina 2006; Brody \textit{et al}. 2007,
2008a,b; Rutkowski \& Yu 2007; Hughston \& Macrina 2008). In this
framework---hereafter referred to as the BHM framework---the price
process of an asset is derived from the specification of (a) future
cash flows associated with the asset, and (b) the flow of
information accessible to market participants. The price is then
given by the discounted risk-adjusted expectation of the cash flows
conditional on the available information.

The simplest model that arises in the BHM framework is briefly
reviewed in \S\ref{sec:2}. In this setup the asset is characterised
by a contract that delivers a single random cash flow at a
predesignated time. Such an asset can be interpreted as having the
structure of a credit-risky discount bond. In \S\ref{sec:3} we
consider the problem of quantifying the amount of information
contained in the bond price concerning the value of the future bond
payout. To this end we determine the mutual information (Yaglom \&
Yaglom 1983, Cover \& Thomas 1999) of these two random variables.
Initially the market has no information, beyond that already
implicit in the asset price, concerning the value of the cash flow.
However, as time goes by, the market gathers information. When the
amount of information reaches the level of the initial entropy of
the cash flow, the market finally `learns' what the value of the
cash flow is. The information-theoretic analysis is extended in
\S\ref{sec:4} where we show that the mutual information at time $t$
is given by the initial uncertainty, less the expected uncertainty
that remains at that time.

In \S\ref{sec:5} a simple model for an informed trader is
introduced. In our approach the informed trader is more susceptive
to the flow of market information than other market participants,
and thus on average is able to estimate the value of the impending
cash flow more quickly and accurately than the other market
participants. Simulation studies show a comparison of the sample
paths for the market price process and the corresponding valuations
made by the informed trader, revealing various intuitive as well as
counterintuitive properties of these processes. The dynamics of the
valuations estimated by the informed trader are worked out in
\S\ref{sec:6}, where we obtain the associated innovations
representation. With a basic model for an informed trader at hand we
are able to quantify the amount of added information held by the
trader. This is worked out in \S\ref{sec:7}, where we construct an
elementary trading strategy making use of the additional
information, demonstrating the existence of statistical arbitrage
opportunities in such circumstances.

\section{Information and asset pricing}
\label{sec:2}

The simplest model arising in the BHM framework can be summarised
thus. We fix a probability space $(\Omega,{\mathcal F},{\mathbb
Q})$, where ${\mathbb Q}$ denotes the risk-neutral measure. We write
${\mathbb E}$ for expectation with respect to ${\mathbb Q}$. The
market is not assumed complete, but we assume the absence of
arbitrage and the existence of an established pricing kernel---these
assumptions ensure the existence of a preferred pricing measure (cf.
Cochrane 2005). We let $X_T$ denote the random cash flow associated
with the asset and paid at time $T$ (for example, the payout of a
credit-risky discount bond). Before time $T$ market participants do
not have direct access to the value of the cash flow. We assume
nevertheless that partial information concerning the value of $X_T$,
obscured by the market noise, can be obtained before time $T$. This
noisy `observation' of $X_T$ generates the market filtration
$\{{\mathcal F}_t\}$, and the price at time $t$ is given by the
risk-neutral expectation of the discounted cash flow, conditional on
${\mathcal F}_t$.

We assume that the `signal' of the noisy observation concerning
$X_T$ is revealed to the market at a constant rate $\sigma$, that
the `noise' is generated by an independent Brownian bridge process
$\{\beta_{tT}\}$, and that the market filtration is generated by an
information process $\{\xi_t\}$ defined for $0\leq t\leq T$ by
\begin{eqnarray}
\xi_t =\sigma t X_T + \beta_{tT}. \label{eq:2.1}
\end{eqnarray}
In other words, $\{{\mathcal F}_t\}=\sigma(\{\xi_s\}_{0 \leq s\leq
t})$. The use of a bridge process for the noise is motivated by the
idea that at time $0$ all the available information about $X_T$ is
incorporated into the \textit{a priori} distribution, and that at
time $T$ the value of $X_T$ is revealed and there is no remaining
noise: the choice of a Brownian bridge is made for simplicity and
tractability. If we further assume that the default-free system of
interest rates is deterministic and let $\{P_{tT}\}$ denote the
price at time $t$ of a discount bond that matures at $T$, then the
price of the credit-risky discount bond at time $t$ is given by
$B_{tT}= P_{tT}{\mathbb E}[X_T|{\mathcal F}_t]$. In the case where
$X_T$ takes the discrete values $\{x_i\}_{ i=1,\ldots,n}$ with the
\textit{a priori} probabilities $\{p_i\}_{i=1,\ldots, n}$, a
calculation shows that
\begin{eqnarray}
B_{tT} = P_{tT} \frac{\sum_{i=1}^n p_i x_i \exp\left[
\frac{T}{T-t}\left(\sigma x_i \xi_t-\frac{1}{2}\sigma^2 x_i^2
t\right)\right]}{\sum_{i=1}^n p_i \exp \left[ \frac{T}{T-t}\left(
\sigma x_i \xi_t-\frac{1}{2} \sigma^2 x_i^2 t \right) \right]}.
\label{eq:2.2}
\end{eqnarray}
This follows from the fact that the conditional risk-neutral
probability defined by $\pi_{it}={\mathbb Q}(X_T=x_i|{\mathcal
F}_t)$ takes the form
\begin{eqnarray}
\pi_{it}=\frac{p_i\exp\left[\frac{T}{T-t}(\sigma x_i \xi_t-
\frac{1}{2} \sigma^2 x_i^2 t)\right]} {\sum_{i=1}^n \,p_i\exp\left[
\frac{T}{T-t}(\sigma x_i \xi_t- \frac{1}{2} \sigma^2 x_i^2
t)\right]}. \label{eq:2.3}
\end{eqnarray}
By taking the stochastic differential of (\ref{eq:2.2}) one finds
that the dynamical equation satisfied by the bond price is
\begin{eqnarray}
\rd B_{tT}=r_t B_{tT}\,\rd t+\frac{\sigma T}{T-t}P_{tT}V_{tT} \, \rd
W_t, \label{eq:2.4}
\end{eqnarray}
where $r_t=-\partial \ln P_{0t}/\partial t$, and where the process
$\{W_t\}$ defined by the expression
\begin{eqnarray}
W_t = \xi_t + \int_0^t \frac{1}{T-s}\,\xi_s\,\rd s - \sigma T
\int_0^t \frac{1}{T-s}\,{\hat X}_{s}\,\rd s \label{eq:2.5}
\end{eqnarray}
turns out to be a standard $\{{\mathcal F}_t\}$-Brownian motion.
Here ${\hat X}_{t}=\sum_{i=1}^n x_i \pi_{it}$ denotes the
conditional expectation of the cash flow $X_T$, and
\begin{eqnarray}
V_{tT}=\sum_{i=1}^n \left(x_i-{\hat X}_{t} \right)^2\pi_{it}
\end{eqnarray}
is the conditional variance of $X_T$ (see Macrina 2006 and Brody
\textit{et al}. 2007 for derivations of the foregoing results).

We observe that in the information-based framework it is possible to
\textit{deduce} the diffusive dynamics (\ref{eq:2.4}) for the price
process, starting from the specification of the cash flow $X_T$ and
the information process $\{\xi_t\}$. The theme that underlies this
framework is that the market acts as a `signal processor' for future
cash flows so as to generate the dynamics of asset prices. This
point of view is natural as a basis for understanding the elements
of price formation, since investment decisions are often made in
accordance with the perceptions of market participants concerning
the future cash flows associated with the given asset.

As far as the market filtration is concerned, the information
contained in $\{\xi_t\}$ is equivalent to that in $\{B_{tT}\}$: we
have $\sigma(\{\xi_s\}_{0\leq s\leq t}) = \sigma(\{B_{sT}\}_{0 \leq
s\leq t})$. This follows from the fact that one can write $B_{tT}=
B(t,\xi_t)$, where
\begin{eqnarray}
B(t,x) = P_{tT} \frac{\sum_{i=1}^n p_i x_i \exp\left[
\frac{T}{T-t}\left(\sigma x_i x-\frac{1}{2}\sigma^2 x_i^2
t\right)\right]}{\sum_{i=1}^n p_i \exp \left[ \frac{T}{T-t}\left(
\sigma x_i x-\frac{1}{2} \sigma^2 x_i^2 t \right) \right]},
\label{eq:2.6}
\end{eqnarray}
from which by differentiation we deduce that
\begin{eqnarray}
B'(t,x) = \frac{\sigma T P_{tT}}{T-t} \frac{\sum_{i=1}^n p_i
\left[x_i-B(t,x)/P_{tT}\right]^2 \exp\left[
\frac{T}{T-t}\left(\sigma x_i x-\frac{1}{2}\sigma^2 x_i^2
t\right)\right]}{\sum_{i=1}^n p_i \exp \left[ \frac{T}{T-t}\left(
\sigma x_i x-\frac{1}{2} \sigma^2 x_i^2 t \right) \right]},
\label{eq:3.9}
\end{eqnarray}
which is positive. Therefore, $B(t,x)$ is monotonically increasing
in $x$, and hence invertible. It follows that from knowledge of the
trajectory $\{\xi_s\}_{0\leq s\leq t}$ one can construct
$\{B_{sT}\}_{0 \leq s\leq t}$; conversely from knowledge of the
trajectory $\{B_{sT}\}_{0 \leq s\leq t}$ one can construct
$\{\xi_s\}_{0\leq s\leq t}$.

\section{Amount of information about the future cash flow
contained in the price process} \label{sec:3}

We would like to quantify how much information regarding the value
of the cash flow $X_T$ is contained in the value at time $t$ of
the information process (\ref{eq:2.1}). A reasonable measure for
such quantification is given by the mutual information $J(\xi_t
,X_T)$ between the two random variables (Shannon \& Weaver 1949;
Gel'fand \textit{et al}. 1956; Gel'fand \& Yaglom 1957), which in
the present context is given by the expression
\begin{eqnarray}
J(\xi_t,X_T) = \sum_{i=1}^n \int_{-\infty}^\infty \rho_{{}_{\xi X}}
(x,i)\, \ln\left( \frac{\rho_{{}_{\xi X}} (x,i)}
{\rho_{{}_{\xi}}(x)\rho_{{}_{X}}(i)}\right) \rd x, \label{eq:3.1}
\end{eqnarray}
where
\begin{eqnarray}
\rho_{{}_{\xi X}}(x,i) = \frac{\rd}{\rd x}\, {\mathbb Q}\Big[
(\xi_t<x) \cap (X_T=x_i)\Big]
\end{eqnarray}
is the joint density function of the random variables $(\xi_t,X_T)$,
and $\rho_{{}_{\xi}}(x)$ and $\rho_{{}_{X}}(i)$ are the respective
marginal probabilities. By use of the relation
\begin{eqnarray}
{\mathbb Q}\Big[ (\xi_t<x) \cap (X_T=x_i)\Big] = {\mathbb Q}(\xi_t<x
| X_T=x_i)\,{\mathbb Q}(X_T=x_i)
\end{eqnarray}
we deduce that
\begin{eqnarray}
\rho_{{}_{\xi X}}(x,i) = p_i \frac{1}{\sqrt{2\pi t(T-t)/T}} \exp
\left(-\half \frac{(x-\sigma x_it)^2}{t(T-t)/T}\right),
\label{eq:2.9}
\end{eqnarray}
since conditional on $X_T=x_i$ the random variable $\xi_t$ is
normally distributed with mean $\sigma x_i t$ and variance
$t(T-t)/T$. From (\ref{eq:2.9}) the marginal densities
\begin{eqnarray}
\rho_{{}_{\xi}}(x) = \sum_{i=1}^n \rho_{{}_{\xi X}}(x,i) \quad {\rm
and} \quad \rho_{{}_{X}}(i) = \int_{-\infty}^\infty \rho_{{}_{\xi
X}}(x,i) \rd x
\end{eqnarray}
can be deduced at once. In particular, we have
$\rho_{{}_{X}}(i)=p_i$, as it should be.

An alternative way of deriving the mutual information in this
context is to make use of the identity
\begin{eqnarray}
J(\xi_t,X_T) = H(\xi_t) - H(\xi_t|X_T). \label{eq:3.6}
\end{eqnarray}
Here $H(\xi_t)=-{\mathbb E}[\ln\rho_\xi(\xi_t)]$ is the entropy of
the random variable $\xi_t$ (Wiener 1948, Khintchine 1953), and
$H(\xi_t|X_T)=-{\mathbb E}[\ln\rho_\xi(\xi_t|X_T)]$ is the entropy
of $\xi_t$ conditional on $X_T$. The conditional density function
$\rho_{{}_{\xi}}(x|X_T)$, $x\in{\mathds R}$, is defined by
$\rho_{{}_{\xi}}(x|X_T)= \rd {\mathbb Q}(\xi_t<x|X_T)/\rd x$. Since
conditional on $X_T$ the random variables $\xi_t$ and $\beta_{tT}$
are both normally distributed, with the same variance $t(T-t)/T$,
and since the entropy of a normally distributed random variable is
independent of its mean, we find that $H(\xi_t|X_T)= H(\beta_{tT})$.
In other words, the mutual information in the present context is
given by the difference of the two entropies:
\begin{eqnarray}
J(\xi_t,X_T) = H(\xi_t) - H(\beta_{tT}). \label{eq:3.7}
\end{eqnarray}
As a consequence, the information about the cash flow $X_T$
contained in $\xi_t$ can be determined (a different approach to
extracting information concerning the asymptotic dividend stream
from option price data is considered in Geman \textit{et al}. 2007).

From an information-theoretic point of view a pair of processes
related through an invertible smooth function, and thus sharing the
\textit{same} filtration, in general possess \textit{different}
information content (entropy). On the other hand, since what is
directly observed in the market is the price $B_{tT}$, which is an
invertible function of $\xi_t$, one might argue that it is more
relevant to determine the mutual information $J(B_{tT},X_T)$, that
is, the amount of information about the future cash flow contained
in the market price. However, since mutual information is given by a
difference of entropies, and since changes in the two entropies
resulting from the transform cancel, we have $J(B_{tT},X_T)=
J(\xi_{t},X_T)$. Therefore, the amount of information about $X_T$
contained in $B_{tT}$ is given by (\ref{eq:3.7}).

\begin{figure}
\begin{center}
  \includegraphics[scale=0.8]{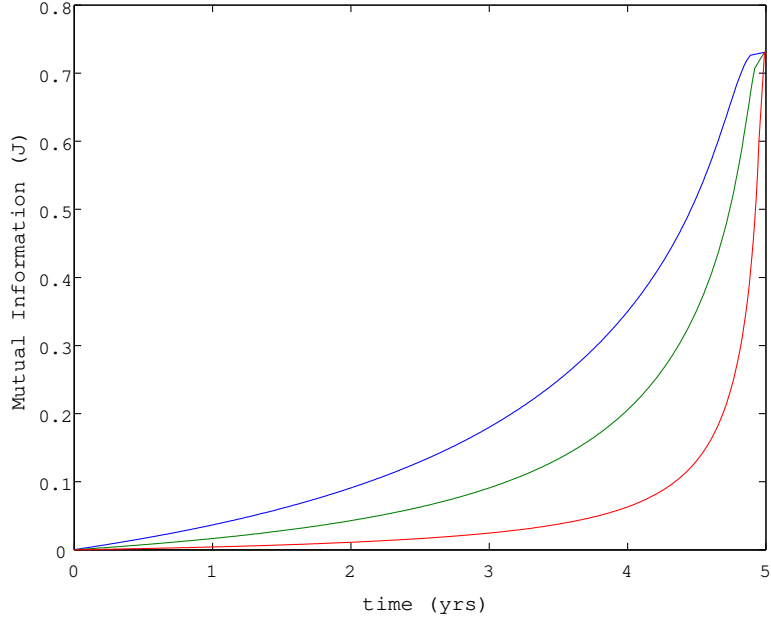}
  \caption{Mutual information $J$ concerning the value of the future
cash flow $X_T$ contained in the price $B_{tT}$. Early observation
of the price yields little information concerning the value of
$X_T$. As time progresses the market extracts more information about
$X_T$; eventually, close to maturity when the value of $X_T$ is to
be revealed the amount of information gained by the market reaches
$-\sum_i p_i \ln p_i$. The parameters are chosen to be $h_1=0$,
$h_2=0.5$, $h_3=1$, $p_1=0.1$, $p_2=0.15$, $p_3=0.75$, and $T=5$.
The three plots correspond to $\sigma=0.75$ (blue), $\sigma=0.5$
(green), and $\sigma=0.25$ (red). The terminal value of $J$ in these
examples is given approximately by $0.73$.
  \label{fig:1}
  }
\end{center}
\end{figure}

In Figure~\ref{fig:1} we plot the mutual information $J(B_{tT},X_T)$
as a function of $t\in[0,T]$ for three values of the information
flow-rate parameter $\sigma$. The information gained by market
participants increases more rapidly as $\sigma$ is raised. On the
other hand, the dynamical relation (\ref{eq:2.4}) shows that the
value of $\sigma$ determines the overall magnitude of the price
volatility. Thus it is possible to quantify the market information
gain and compare this with the price volatility.

To see how entropy transforms under a nonlinear invertible map,
suppose that $X$ is a random variable with density $p(x)$, and that
$Y$ is another random variable given by $Y=f(X)$, where $f(x)$ is
smooth and invertible. Then the density function for $Y$ is given by
$q(y)= p(f^{-1}(y)) / f'(f^{-1}(y))$. Substituting this in
$H(Y)=-\int q(y) \ln q(y) \rd y$ and changing variables by setting
$y=f(x)$, we find that $H(Y) = H(X) + \int p(x) \ln f'(x) \rd x$. As
a consequence, we have
\begin{eqnarray}
H(B_{tT}) = H(\xi_t) + \int_{-\infty}^\infty \rho_\xi(x) \ln
B'(t,x)\, \rd x, \label{eq:3.8}
\end{eqnarray}
where $B'(t,x)=\partial B(t,x)/\partial x$. The advantage of this
expression is that we need not determine the inverse of the function
$B(t,x)$ defined in (\ref{eq:2.6}) in order to calculate
$H(B_{tT})$. From (\ref{eq:3.9}) it follows that
\begin{eqnarray}
\int_{-\infty}^\infty \rho_\xi(x) \ln B'(t,x)\, \rd x = {\mathbb E}
\left[ \ln \left( \frac{\sigma T}{T-t}\, P_{tT}  V_{tT} \right)
\right].
\end{eqnarray}
In other words, the difference between $H(B_{tT}\!)$ and $H(\xi_t)$
is the average log-volatility of the price process at time $t$.

\section{Analysis of information measures}
\label{sec:4}

We proceed in this section to consider the Shannon-Wiener entropy
associated with the conditional risk-neutral probabilities. Analysis
of the entropy leads to insights into the qualitative behaviour of
the asset price volatility. The Shannon-Wiener entropy is defined by
the expression
\begin{eqnarray}
H_t = -\sum_{i=1}^n \pi_{it} \ln \pi_{it}. \label{eq:4.1}
\end{eqnarray}
We shall demonstrate that the mutual information (\ref{eq:3.1}) and
the Shannon-Wiener entropy (\ref{eq:4.1}) are related as follows:
\begin{eqnarray}
J(\xi_{t},X_T)=H_0-{\mathbb E}[H_t]. \label{eq:4.2}
\end{eqnarray}
Thus, the mutual information at time $t$ is given by the initial
uncertainty, less the expected uncertainty that still remains at
that time.

The derivation of (\ref{eq:4.2}) proceeds in two steps. First we
shall show that
\begin{eqnarray}
{\mathbb E}[H_t] = H_0 - \half\, {\mathbb E} \left[ \int_0^t
\frac{\sigma^2 T^2}{(T-s)^2}\, V_{sT} \rd s \right], \label{eq:4.3}
\end{eqnarray}
and then we shall show that
\begin{eqnarray}
J(\xi_{t},X_T) = \half {\mathbb E} \left[ \int_0^t \frac{\sigma^2
T^2}{(T-s)^2}\, V_{sT} \rd s \right]. \label{eq:4.4}
\end{eqnarray}

\textit{Proof of (\ref{eq:4.3})}. Let us begin by deriving the
dynamical equation satisfied by the Shannon-Wiener entropy. From
(\ref{eq:2.3}) and (\ref{eq:2.5}) we find that the conditional
probability satisfies
\begin{eqnarray}
\rd\pi_{it} = \frac{\sigma T}{T-t}\, (x_i-{\hat X}_t)\pi_{it}\rd
W_t. \label{eq:4.5}
\end{eqnarray}
It follows, by an application of Ito's lemma to (\ref{eq:4.1}), that
\begin{eqnarray}
H_{t} = H_0 - \half \int_0^t \frac{\sigma^2 T^2}{(T-s)^2}\, V_{st}
\rd s - \int_0^t \frac{\sigma T}{T-s} \left({\textstyle
\sum\limits_{i=1}^n} (x_i-{\hat X}_s) \pi_{is} \ln \pi_{is}\right)
\rd W_s. \label{eq:4.6}
\end{eqnarray}
Taking the expectation of both sides of (\ref{eq:4.6}), we obtain
(\ref{eq:4.3}), as desired. \hspace*{\fill} $\square$ \vspace{0.2cm}

\textit{Proof of (\ref{eq:4.4})}. In Gel'fand \& Yaglom (1957) it is
shown that the mutual information can be expressed as the
expectation of the log density of the joint measure $\mu_{{}_{\xi
X}}$ with respect to the product measure $\mu_{{}_{\xi}}
\!*\!\mu_{{}_X}$:
\begin{eqnarray}
J(\xi_t,X_T) = {\mathbb E}\left[ \ln \frac{\rd \mu_{{}_{\xi X}}}
{\rd(\mu_{{}_{\xi}}\!*\!\mu_{{}_X})} \right]. \label{eq:4.7}
\end{eqnarray}
We are thus required to determine the relevant Radon-Nikodym
derivative. We shall follow the line of argument presented in Davis
(1978) (see, also, Duncan 1970). To proceed, we require the
introduction of an auxiliary measure ${\mathbb B}$ introduced in
Macrina (2006) and Brody \textit{et al}. (2007, 2008a). This is the
so-called bridge measure under which the information process
$\{\xi_t\}$ becomes a Brownian bridge. The argument goes as follows.
We fix the probability space $(\Omega, {\mathcal F},{\mathbb Q})$
with a filtration $\{{\mathcal H}_t \}_{0\leq t<\infty}$, and
introduce a ${\mathbb Q}$-Brownian motion $\{B_t\}$ such that the
Brownian bridge $\{\beta_{tT}\}$ appearing in (\ref{eq:2.1}) is
given by
\begin{eqnarray}
\beta_{tT} = (T-t)\int_0^t \frac{1}{T-s}\, \rd B_s. \label{eq:4.8}
\end{eqnarray}
This is the standard integral representation for a Brownian bridge
(see, e.g., Hida 1980, Protter 2005). Setting $\nu_t=\sigma T/(T-t)$
we define
\begin{eqnarray}
\Lambda_t^{-1} = \exp\left( - \int_0^t \nu_s X_T \rd B_s - \half
\int_0^t \nu_s^2 X_T^2 \rd s \right), \label{eq:4.9}
\end{eqnarray}
where $X_T$ is ${\mathbb Q}$-independent of $\{B_t\}$ and is
${\mathcal H}_0$-measurable. For fixed $u<T$ we introduce the
measure ${\mathbb B}$ on ${\mathcal H}_u$ by writing
\begin{eqnarray}
\rd {\mathbb B} = \Lambda_u^{-1} \rd {\mathbb Q}. \label{eq:4.10}
\end{eqnarray}
Then the process $\{W_t^*\}_{0\leq t\leq u}$ defined by
\begin{eqnarray}
W_t^* = \int_0^t \nu_s X_T \rd s + B_t \label{eq:4.11}
\end{eqnarray}
is a ${\mathbb B}$-Brownian motion, since $\nu_s X_T$ is bounded for
any $s\leq u$.

Under ${\mathbb B}$ we find that the distribution of $X_T$ is same
as it is under ${\mathbb Q}$, that $\{\xi_t\}$ and $X_T$ are
independent, and that $\{\xi_t\}$ is a ${\mathbb B}$-Brownian bridge
(cf. Brody \textit{et al}. 2008a). To see these, recall that since
$X_T$ and $\{\beta_{tT}\}$, and hence $X_T$ and $\{B_t\}$, are
independent, the probability law of $\{B_t\}$ conditional on $X_T$
remains that of a Brownian motion. Now take a bounded function
$f(x)$ and consider
\begin{eqnarray}
{\mathbb E}^{\mathbb B}\left[ f(X_T)\right] = {\mathbb E} \left[
f(X_T){\mathbb E} [\Lambda_u^{-1}|X_T]\right]. \label{eq:4.12}
\end{eqnarray}
Conditional on $X_T$, $\Lambda_u^{-1}$ takes the form of a Girsanov
exponential, since $\{B_t\}$ is a ${\mathbb Q}$-Brownian motion.
Therefore, the inner expectation equals unity, and we find
\begin{eqnarray}
{\mathbb E}^{\mathbb B}\left[ f(X_T)\right]={\mathbb E} \left[
f(X_T)\right] \label{eq:4.13}
\end{eqnarray}
for every bounded function $f(x)$. In other words, $X_T$ has the
same probability law under ${\mathbb Q}$ and ${\mathbb B}$. In a
similar manner, for any sequence of times $t_1,\ldots,t_n \in [0,u]$
and any bounded function $g: {\mathds R}^n\to{\mathds R}$ we wish to
calculate
\begin{eqnarray}
{\mathbb E}^{\mathbb B}\left[ f(X_T) g(\xi_{t_1},\cdots,\xi_{t_n})
\right] = {\mathbb E} \Big[ f(X_T){\mathbb E} \left[ g(\xi_{t_1},
\cdots, \xi_{t_n}) \Lambda_u^{-1}|X_T \right]\Big] . \label{eq:4.14}
\end{eqnarray}
By the same argument as above, for each $X_T$ the process
$\{W_t^*|_{X_T}\}_{0\leq t\leq u}$ is Brownian under the measure
whose density is $\Lambda_u^{-1}|_{X_T}$. Since $\{W_t^*\}_{0\leq
t\leq u}$ itself is Brownian under ${\mathbb B}$ we have
\begin{eqnarray}
{\mathbb E} \left[g(\xi_{t_1}, \cdots, \xi_{t_n}) \Lambda_u^{-1}|X_T
\right] = {\mathbb E}^{\mathbb B}\left[ g(\xi_{t_1}, \cdots,
\xi_{t_n}) \right], \label{eq:4.15}
\end{eqnarray}
and hence
\begin{eqnarray}
{\mathbb E}^{\mathbb B}\left[ f(X_T) g(\xi_{t_1},\cdots,\xi_{t_n})
\right] = {\mathbb E} \left[ f(X_T)\right] {\mathbb E}^{\mathbb
B}\left[g(\xi_{t_1},\cdots,\xi_{t_n})\right]. \label{eq:4.16}
\end{eqnarray}
However, ${\mathbb B}$ and ${\mathbb Q}$ coincide on $X_T$ so that
${\mathbb E}^{\mathbb B}[f(X_T)]={\mathbb E}[f(X_T)]$, from which it
follows that $X_T$ and $\{\xi_t\}$ are ${\mathbb B}$-independent. By
combining (\ref{eq:4.8}) and (\ref{eq:4.11}) we get
\begin{eqnarray}
\rd W_t^* = \nu_t X_T \rd t + \frac{1}{T-t}\,\beta_{tT}\rd t + \rd
\beta_{tT}. \label{eq:4.17}
\end{eqnarray}
Eliminating $\beta_{tT}$ by use of $\beta_{tT}=\xi_t-\sigma t X_T$
we obtain the relation
\begin{eqnarray}
\rd \xi_t = -\frac{1}{T-t}\,\xi_t\rd t + \rd W_t^*, \label{eq:4.18}
\end{eqnarray}
which is the dynamical equation satisfied by a Brownian bridge in
the ${\mathbb B}$ measure.

Let $\Psi$ be the map: $\omega \to \{\{\xi_t(\omega)\}_{0\leq t\leq
u},X_T(\omega)\}$. Then the joint sample space measure of
$\{\{\xi_t(\omega)\},X_T(\omega)\}$ is $\mu_{{}_{\xi X}}
(A)={\mathbb Q}(\Psi^{-1}(A))$ for any measurable set $A$, and the
sample space measure of $X_T$ is $\mu_{{}_X}(A')={\mathbb Q}(
X_T^{-1}(A'))$ for any measurable set $A'$. However, since
$\{\xi_t\}$ and $X_T$ are independent under ${\mathbb B}$, and
$\{\xi_t\}$ is a ${\mathbb B}$-Brownian bridge, we have
$\mu_{{}_X}\!*\!\mu_{{}_\beta}(A)={\mathbb B}(\Psi^{-1}(A))$. It
follows from (\ref{eq:4.9}) that
\begin{eqnarray}
\frac{\rd{\mathbb Q}}{\rd{\mathbb B}} = \exp\left(\int_0^u \nu_s X_T
\rd B_s + \half \int_0^u \nu_s^2 X_T^2 \rd s \right).
\label{eq:4.19}
\end{eqnarray}
Substituting (\ref{eq:4.11}) in here we thus deduce that
\begin{eqnarray}
\frac{\rd\mu_{{}_{\xi X}}} {\rd(\mu_{{}_X}\!*\!\mu_{{}_\beta})} =
\exp \left( \int_0^u \nu_s X_T \rd W_s^* - \half \int_0^u \nu_s^2
X_T^2 \rd s \right). \label{eq:4.20}
\end{eqnarray}

Turning to the innovations representation (\ref{eq:2.5}) we find,
along with (\ref{eq:4.18}), that $\{W_t\}$ and $\{W_t^*\}$ are
related according to
\begin{eqnarray}
\rd W_t^* = \nu_t {\hat X}_t \rd t + \rd W_t. \label{eq:4.21}
\end{eqnarray}
Thus, following a similar line of argument we obtain
\begin{eqnarray}
\frac{\rd\mu_{{}_\xi}}{\rd\mu_{{}_\beta}} = \exp \left( \int_0^u
\nu_s {\hat X}_s\rd W_s^* - \half \int_0^u \nu_s^2 {\hat X}_s^2 \rd
s \right), \label{eq:4.22}
\end{eqnarray}
which is a version of the likelihood ratio formula of Kailath
(1971). The measure $\mu_{{}_\xi}$ thus corresponds to the `signal
present' situation, while the `signal absent' case corresponds to
$\{\xi_t\}$ being pure bridge noise with measure $\mu_{{}_\beta}$.
Combining (\ref{eq:4.20}) and (\ref{eq:4.22}), and making use of
(\ref{eq:4.11}), we deduce
\begin{eqnarray}
\frac{\rd\mu_{{}_{\xi X}}}{\rd(\mu_{{}_\xi}\!*\!\mu_{{}_X})} = \exp
\left( \int_0^u \nu_s (X_T-{\hat X}_s)\rd B_s + \half \int_0^u
\nu_s^2 (X_T-{\hat X}_s)^2 \rd s \right). \label{eq:4.23}
\end{eqnarray}
Taking the expectation of the logarithm of this, bearing in mind
that $\{B_t\}$ is a ${\mathbb Q}$-Brownian motion, we recover
(\ref{eq:4.4}), as claimed. \hspace*{\fill} $\square$ \vspace{0.2cm}

The entropy process $\{H_t\}_{0\leq t<T}$ has the property that
$\lim_{t\to T}H_t=0$. This follows from the fact that the
conditional probability process $\{\pi_{it}\}_{0\leq t<T}$ has the
limiting behaviour
\begin{eqnarray}
\lim_{t\to T} \pi_{it}(\omega) = {\mathds 1}{\{X_T(\omega)=x_i\}}
\label{eq:4.24}
\end{eqnarray}
for $i=1,\ldots,n$. The proof of (\ref{eq:4.24}) is as follows. Let
$\omega\in\Omega$ be fixed, and suppose that $X_T(\omega)=x_k$ for
some $k$. For this realisation of $\omega$ the information process
is given by $\xi_t=\sigma t x_k+\beta_{tT}$. Substituting this
expression for $\xi_t$ into (\ref{eq:2.3}) and dividing the
denominator and the numerator by the exponential factor appearing in
the numerator, we deduce that
\begin{eqnarray}
\pi_{kt} = \frac{p_k}{p_k+\sum\limits_{j\neq k} p_j \exp\left[
\frac{T}{T-t}\left( \sigma(x_j-x_k)\beta_{tT}-\frac{1}{2} \sigma^2 t
(x_j-x_k)^2\right)\right]}. \label{eq:4.25}
\end{eqnarray}
Observe that all of the terms in the sum in the denominator vanish
as $t$ approaches $T$, and therefore $\lim_{t\to T} \pi_{kt}=1$. It
follows that $\lim_{t\to T}\pi_{it}=0$ for $i\neq k$, and thus
(\ref{eq:4.24}). Finally, since $H_t = -\ln \prod_{i=1}^n
\pi_{it}^{\pi_{it}}$ by (\ref{eq:4.1}) and since $\lim_{t\to T}
\pi_{it}^{\pi_{it}}=1$, we deduce that $\lim_{t\to T}H_t=0$.

If we let $t$ approach $T$ in (\ref{eq:4.3}) we find the following
relation:
\begin{eqnarray}
H_0 = \half {\mathbb E} \left[ \int_0^T \frac{\sigma^2
T^2}{(T-s)^2}\, V_{sT} \rd s \right]. \label{eq:4.26}
\end{eqnarray}
Since $H_0$ is bounded by $\ln n$, where $n$ is the number of values
$X_T$ can take, and since the coefficient of the conditional
variance $\{V_{sT}\}$ in the integrand diverges quadratically as $s$
approaches $T$, this relation shows that the variance process has to
decay sufficiently rapidly to ensure the existence of the right side
of (\ref{eq:4.26}). On the other hand, the conditional variance also
generates the random movement of the asset price volatility in
(\ref{eq:2.4}). As a consequence we are able to obtain a crude
estimate of the magnitude of the cumulative volatility. It is worth
noting that in models based on Brownian noise the entropy and mutual
information are closely related to prices of variance or volatility
derivatives. A related observation has been made by Soklakov (2008).

It should be remarked that the limiting behaviour $\lim_{t\to
T}H_t=0$ is specific to the case in which $X_T$ takes discrete
values. If $X_T$ is a continuous random variable, then the
associated entropy has the property that $\lim_{t\to T}H_t=-\infty$,
which can be seen from the Hirschman inequality in Fourier analysis
(Beckner 1975):
\begin{eqnarray}
H_t \leq \half \left( 1 + \ln(2\pi)\right) + \half \ln V_{tT}.
\label{eq:4.27}
\end{eqnarray}
It follows from (\ref{eq:4.6}) that the variance process
$\{V_{sT}\}$ in this case does not vanish sufficiently rapidly to
ensure finiteness of the right side of (\ref{eq:4.3}) as $t$
approaches $T$. In other words, there is a qualitative difference in
the behaviour of the volatility process, depending on whether the
cash flow is a continuous or discrete random variable. In
particular, volatility products may be overpriced in models based on
continuous cash flows, since real market cash flows are not
continuous.

\section{A model for an informed trader}
\label{sec:5}

In the previous sections we have examined the BHM framework from an
information theoretic perspective. In particular, we have been able
to quantify the amount of information about the future cash flow of
an asset contained in its price. We turn now to consider a model for
an informed trader who has access to an additional information
source, apart from the price process itself, concerning the future
return of an asset. We assume that the informed trader is `small' in
the sense that access to the additional information is limited, and
that the actions of the informed trader will not impact the price
process. In other words, the model is not for a large number of
small agents; rather, it is for a single agent, or a highly
restricted number of agents, who carefully execute their trading
strategies, taking advantage of the additional information.

One might expect that the use of additional information gives a
\textit{definite} advantage for the trader. This, however, is not
necessarily the case: additional information is in general obscured
by additional noise. As a consequence, the valuation process
estimated by the informed trader can entail higher volatility than
the actual market price movements. It follows that any strategy
making use of additional information will tend to embody additional
risk. Nevertheless, on average such strategies are expected to
outperform the market, and this is the idea behind some of the
statistical arbitrage strategies adopted by hedge funds.

The BHM framework is sufficiently flexible to model this kind of
scenario. Indeed, the use of this framework as a basis for the
development of insider-trading models was recognised early on
(Macrina 2006). Our intention here is to apply such ideas to
describe the disparity in the ability of processing publicly
available information, and to illustrate how statistical arbitrage
opportunities can be seen to arise in a simple model. It should be
emphasised that many of the simplifying assumptions---that the asset
entails a single cash flow; that the information flow rates are
constants; that the interest rate is deterministic; and that the
noises are modelled by Brownian bridges---can be relaxed without
affecting the main qualitative features of the model.

\begin{figure}
\begin{center}
  \includegraphics[scale=0.9]{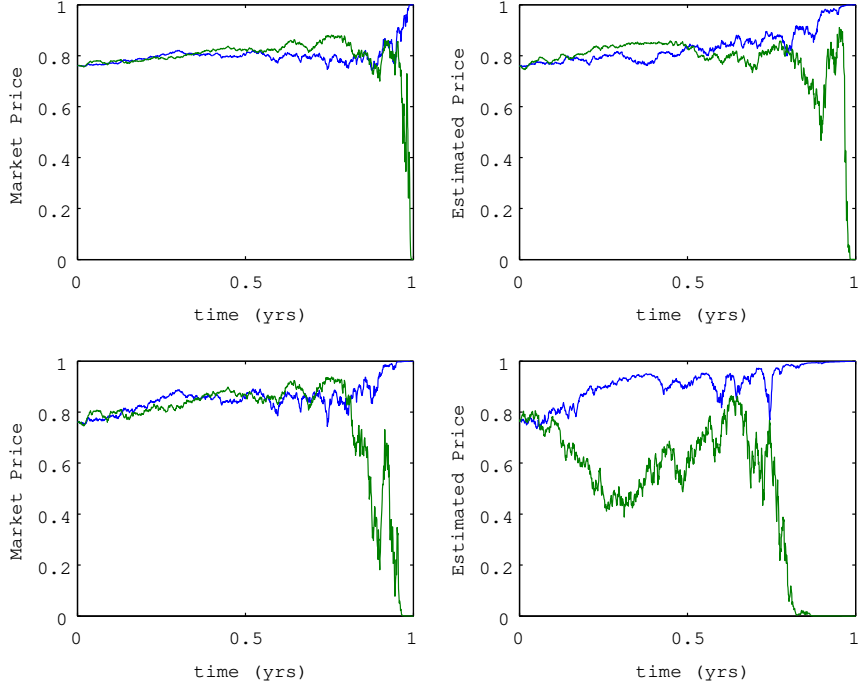}
  \caption{Sample paths comparison for a defaultable digital bond.
Each plot contains a pair of sample paths: one for which the bond
does not default and one for which the bond defaults. The two plots
on the left show the market prices with $\sigma=0.25$ (top) and
$\sigma=0.7$ (bottom). Market prices can be compared with the
valuations estimated by the informed trader, represented by the two
plots on the right. The parameters are chosen to be $\sigma'=0.4$,
$\rho=0.1$ (top) and $\sigma'=0.01$, $\rho=0.9$ (bottom). In all
plots, the other parameters are set to be $h_1=0$, $h_2=1$,
$p_0=0.2$, $p_1=0.8$, $r=0.05$, and $T=1$.
  \label{fig:2}
  }
\end{center}
\end{figure}

We assume the setup for the market outlined in \S\ref{sec:2}.
However, in addition there is an informed trader who has a further
noisy information source represented by the information process
\begin{eqnarray}
\xi_t^\prime = \sigma' t X_T + \beta_{tT}^\prime . \label{eq:5.1}
\end{eqnarray}
The noise term $\{\beta_{tT}^\prime\}$ may or may not be correlated
with the market noise $\{\beta_{tT}\}$. We let $\{B_t\}$ and
$\{B'_t\}$ be a pair of Brownian motions with correlation $\rho$,
and define the associated Brownian bridges by
\begin{eqnarray}
\beta_{tT}=B_t-\frac{t}{T}\,B_T, \quad {\rm and} \quad
\beta'_{tT}=B'_t-\frac{t}{T}\,B'_T . \label{eq:5.2}
\end{eqnarray}
In this way we can model the two noise terms with fair amount of
generality (we may use alternatively the integral representation
(\ref{eq:4.8}) to construct the bridge processes; however, the
choice (\ref{eq:5.2}) is more suitable for simulation purposes),
since $\rho$ determines the correlation of $\{\beta_{tT}\}$ and
$\{\beta'_{tT}\}$. In particular, if $|\rho|=1$ then the informed
trader has two linear equations (\ref{eq:2.1}) and (\ref{eq:5.1})
for the two unknowns $X_T$ and $\beta_{tT}$; hence the value of the
future cash flow $X_T$ will become instantly accessible to the
trader (assuming $|\sigma|\neq|\sigma'|$). This situation
corresponds to the fully-informed conventional `insider' often
considered in the literature. The other extreme, for which
$|\rho|\ll1$, is of interest, since the informed trader must choose
a strategy optimally so as not to be overwhelmed by the additional
noise.

We let $\{{\mathcal F}'_t\}$ denote the filtration generated by
$\{\xi_t^\prime\}$. If $\sigma'>\sigma$, then knowledge of the value
of $X_T$ is revealed at a faster rate in the `primed' filtration.
This, however, does not mean that $\{{\mathcal F}_t\}$ is contained
in $\{{\mathcal F}'_t\}$ even if $\rho=1$; the two filtrations are
merely inequivalent. On the other hand, since the informed trader
also has access to the price process, which is adapted to
$\{{\mathcal F}_t\}$, it is reasonable to assume that the
information source is given by $\{{\mathcal G}_t\}= \sigma(
\{\xi_s,\xi'_s\}_{0\leq s\leq t})$. We assume that the additional
information commences at time $t=0$; hence the \textit{a priori}
probabilities $\{p_i\}$ for $X_T$ to take the values $\{x_i\}$
remain the same. This assumption may seem limiting; however, it is
not unreasonable since the `lifetime' of the extra information, i.e.
the period over which extra information has value, is often short in
practice.

\begin{figure}
\begin{center}
  \includegraphics[scale=0.8]{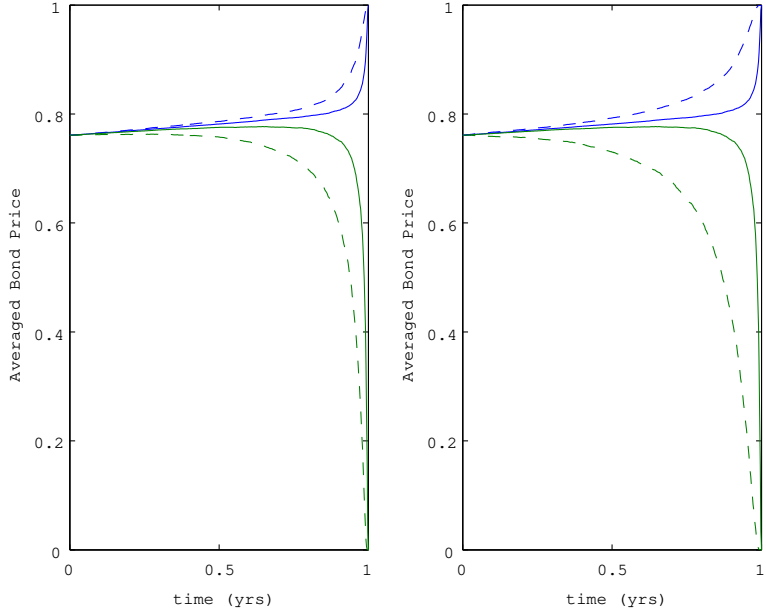}
  \caption{Comparison between averaged market prices and
informed valuations. In both plots, the solid lines represent
averaged market prices of a defaultable digital bond; one which
default does not occur and one which the bond defaulted. The
parameters are chosen so that $X_T$ takes values $0$ or $1$ with
\textit{a priori} default probability $20\%$, $r=5\%$, $T=1$, and
the market signal-to-noise ratio is $\sigma=0.2$. These are compared
with the corresponding averages for the informed trader (dashed
lines), with two sets of parameters: $\sigma'=0.4$ and $\rho=0.1$
(left); and $\sigma'=0$ and $\rho=0.95$ (right).
  \label{fig:3}
  }
\end{center}
\end{figure}

The informed trader will use the extra information to work out the
price that the market would have made had $\{{\mathcal G}_t\}$ been
accessible to general market participants. We shall now calculate
the valuation process made by the informed trader on this basis.
From the Markov property of the joint information process
$\{\xi_t,\xi'_t\}$ we find that the informed valuation process is
given by
\begin{eqnarray}
{\tilde B}_{tT} = P_{tT} \sum_{i=1}^n x_i {\tilde\pi}_{it},
\end{eqnarray}
where ${\tilde\pi}_{it}={\mathbb Q}(X_T=x_i|\xi_t,\xi'_t)$. By use
of the Bayes formula
\begin{equation}
{\tilde\pi}_{it} = \frac{p_i\rho(\xi_t,\xi'_t|X_T=x_i)}
{\sum_{i=0}^n p_i \rho(\xi_t,\xi'_t|X_T=x_i)}, \label{eq:5.4}
\end{equation}
where the conditional density is given by the bivariate normal
density function
\begin{eqnarray}
&& \hspace{-0.7cm} \rho(\xi_t,\xi'_t|X_T=x_i)=\frac{T}{2\pi
t(T-t)\sqrt{1-\rho^2}}\, \exp\left[-\frac{T}{2(1-\rho^2)} \right.
\nonumber \\ && \hspace{-0.2cm} \times  \left. \left(
\frac{(\xi_t-\sigma x_it)^2}{t(T-t)} - \frac{2\rho(\xi_t-\sigma
x_it) (\xi'_t-\sigma' x_i t)}{t(T-t)} + \frac{(\xi'_t-\sigma' x_i
t)^2}{t(T-t)} \right) \right], \label{eq:5.5}
\end{eqnarray}
we deduce that
\begin{eqnarray}
{\tilde B}_{tT}=P_{tT}\frac{\sum_{i=1}^n p_i x_i \exp\left[
\frac{T}{\varrho(T-t)} \left(x_{i}(\sigma_1\xi_t+\sigma_2\xi'_t)-
\frac{1}{2} \sigma_3^2 x_i^2t\right)\right]} {\sum_{i=1}^n p_i \exp
\left[\frac{T}{\varrho(T-t)}\left(x_{i}(\sigma_1\xi_t+\sigma_2\xi'_t)-
\frac{1}{2} \sigma_3^2 x_i^2t\right)\right]}. \label{eq:5.6}
\end{eqnarray}
Here we set $\sigma_1=\sigma-\rho \sigma'$,
$\sigma_2=\sigma'-\rho\sigma$, $\sigma_3^2= \sigma^2
-2\rho\sigma\sigma'+\sigma^{\prime2}$, and $\varrho=1-\rho^2$.

The process (\ref{eq:5.6}) is our model for the valuations made by
the informed trader. It is straightforward to simulate the informed
valuation process $\{{\tilde B}_{tT}\}$ and compare this against the
uninformed market price process $\{B_{tT}\}$. By suitably adjusting
the values for $\sigma$, $\sigma^\prime$, and $\rho$ we are able to
confirm various intuitive aspects of the behaviour of these
processes; some examples are displayed in Figure~\ref{fig:2}. With
respect to any given sample path the expected price of the informed
trader at time $t\in[0,T]$ may be less accurate (by `accurate' we
mean close to the true value) than the market price. However, on
average the valuation determined by the informed trader converges
more rapidly to the true value of the bond. This is illustrated in
Figure~\ref{fig:3} where we plot the averaged sample paths
conditional on the given outcome.

The plot on the right side of Figure~\ref{fig:3} indicates that the
performance of the informed trader is high even if the
signal-to-noise ratio $\sigma'$ of the additional information source
is set to zero (and hence $\{\xi_t'\}$ is pure noise). In fact the
quality of the estimate decreases as the value of $\sigma'$ is
raised from zero, until it reaches the critical level
$\sigma'=\rho\sigma$. Putting the matter differently, \textit{the
quality of the estimate made by an informed trader is not monotonic
in the signal-to-noise ratio of the additional information source}.
This feature may appear counterintuitive, but it can be understood
by rearrangement of terms in (\ref{eq:5.6}) into a form analogous to
(\ref{eq:4.5}):
\begin{eqnarray}
{\tilde B}_{tT}=P_{tT}\frac{p_kx_k+\sum\limits_{i\neq k} p_i x_i
\exp \left[ \frac{T}{\varrho(T-t)} \left(
(1-\rho)(\sigma+\sigma')\omega_{ik}\beta_{tT}-\frac{1}{2} \sigma_3^2
\omega_{ik}^2 t \right) \right]}{p_k+\sum\limits_{i\neq k} p_i \exp
\left[ \frac{T}{\varrho(T-t)} \left(
(1-\rho)(\sigma+\sigma')\omega_{ik}\beta_{tT}-\frac{1}{2} \sigma_3^2
\omega_{ik}^2 t \right) \right]}. \label{eq:5.7}
\end{eqnarray}
Here we write $\omega_{ik}=x_i-x_k$. This expression shows that the
exponential rate of convergence for the process $\{{\tilde
B}_{tT}\}$ to approach the `true' value $x_k$ is governed by the
ratio $\sigma_3^2 tT/[2\varrho(T-t)]$. In particular, for fixed
$\sigma$ and $\rho$ this ratio takes a minimum value at
$\sigma'=\rho\sigma$. When $\sigma'= \rho\sigma$, the linear
equations $\xi_t=\sigma t X_T + \beta_{tT}$ and $\xi'_t=\sigma' t
X_T + \beta'_{tT}$ become closest to being degenerate, and hence the
value of the additional information is minimised.

\section{Innovations and the dynamics of informed valuations}
\label{sec:6}

Our objective now is to obtain an innovations representation for the
valuations made by the informed trader. For this purpose it suffices
to derive the dynamical equation satisfied by the `insider'
valuation $\{{\tilde B}_{tT}\}$, or equivalently, by the conditional
probability $\{{\tilde \pi}_{it}\}$. The calculation simplifies if
we express (\ref{eq:5.6}) in the form
\begin{eqnarray}
{\tilde B}_{tT}=P_{tT}\frac{\sum_{i=1}^n p_i x_i \exp\left[
\frac{T}{T-t} \left({\hat\sigma}x_{i}{\hat\xi}_t-\frac{1}{2}
{\hat\sigma}^2 x_i^2t\right)\right]} {\sum_{i=1}^n p_i \exp
\left[\frac{T}{T-t}\left({\hat\sigma}x_{i}{\hat\xi}_t- \frac{1}{2}
{\hat\sigma}^2 x_i^2t\right)\right]}. \label{eq:6.1}
\end{eqnarray}
Here the process
\begin{eqnarray}
{\hat\xi}_t = {\hat\sigma} t X_T + {\hat\beta}_{tT}, \qquad
{\hat\sigma}^2=\frac{\sigma^2-2\rho\sigma\sigma'+
\sigma^{\prime2}}{1-\rho^2} \label{eq:6.2}
\end{eqnarray}
represents the `enhanced' information being effectively received by
the informed trader, with the modified bridge noise
\begin{eqnarray}
{\hat\beta}_{tT} = \sqrt{\frac{1-\rho^2}{\sigma^2-2\rho
\sigma\sigma'+ \sigma^{\prime2}}} \left[ \frac{\sigma-\rho
\sigma'}{1-\rho^2}\, \beta_{tT} + \frac{\sigma'-\rho
\sigma}{1-\rho^2} \,\beta'_{tT} \right]. \label{eq:6.3}
\end{eqnarray}
Applying Ito's lemma to (\ref{eq:5.4}) and making use of
(\ref{eq:6.2}), we find that
\begin{eqnarray}
\frac{\rd{\tilde\pi}_{it}}{{\tilde\pi}_{it}} =
\frac{{\hat\sigma}T}{T-t}\, (x_i-{\tilde X}_t)\, \rd Z_t,
\label{eq:6.4}
\end{eqnarray}
where ${\tilde X}_t = {\mathbb E}[X_T|{\mathcal G}_t]$. The process
$\{Z_t\}$ appearing in (\ref{eq:6.4}) is defined by
\begin{eqnarray}
Z_t = {\hat\xi}_t + \int_0^t \frac{1}{T-s}\, {\hat\xi}_s \rd s -
\int_0^t \frac{{\hat\sigma}T}{T-s}\, {\tilde X}_s \rd s.
\label{eq:6.5}
\end{eqnarray}
By following a line of argument similar to that presented in Brody
\textit{et al}. (2007) it can be shown that $\{Z_t\}$ is a standard
${\mathcal G}_t$-Brownian motion. It follows that the valuation
process of the informed trader obeys the following dynamical
equation:
\begin{eqnarray}
\rd {\tilde B}_{tT}=r_t {\tilde B}_{tT}\rd t+ \Gamma_t \rd Z_t,
\label{eq:6.6}
\end{eqnarray}
where the volatility process $\{\Gamma_t\}$ is given by
\begin{eqnarray}
\Gamma_t = \frac{{\hat\sigma}T}{T-t}\,P_{tT}{\tilde V}_{tT},
\label{eq:6.7}
\end{eqnarray}
and ${\tilde V}_{tT}$ denotes the variance of $X_T$ conditional on
${\mathcal G}_t$.

The fact that the information $\{\xi_t,\xi'_t\}$ accessible to the
informed trader can be `compactified' into a single enhanced
information $\{{\hat\xi}_t\}$ can be understood as follows. Since
the noise processes $\{\beta_{tT}\}$ and $\{\beta'_{tT}\}$ have
correlation $\rho$, one can write
\begin{eqnarray}
\beta'_{tT}=\rho\beta_{tT}+ \sqrt{1-\rho^2}\,{\bar\beta}_{tT},
\end{eqnarray}
where the Brownian bridge process $\{{\bar\beta}_{tT}\}$ is taken to
be independent of $\{\beta_{tT}\}$. Similarly, we can decompose the
extra information $\{\xi'_t\}$ in the form
\begin{eqnarray}
\xi'_{t}=\rho\xi_{t}+ \sqrt{1-\rho^2}\,{\bar\xi}_{t},
\end{eqnarray}
where ${\bar\xi}_{t}={\bar\sigma}tX_T+{\bar\beta}_{tT}$ and
${\bar\sigma}=(\sigma'-\rho\sigma)/\sqrt{1-\rho^2}$. It should be
evident that the filtration generated jointly by $\{\xi_t,\xi'_t\}$
is equivalent to that generated jointly by $\{\xi_t,{\bar\xi}_t\}$.
However, the information processes $\{\xi_t\}$ and $\{{\bar\xi}_t\}$
have independent noises. To proceed we note that the process
$\{\delta_t\}$ defined by
\begin{eqnarray}
\delta_t = \frac{\xi_t}{\sigma}-\frac{{\bar\xi}_t}{{\bar\sigma}} =
\frac{\beta_{tT}}{\sigma}-\frac{{\bar\beta}_{tT}}{{\bar\sigma}}
\end{eqnarray}
is purely noise, and is independent of $X_T$. We can construct a new
Brownian bridge that is independent of this noise. A standard
orthogonalisation shows that this is given by the bridge process
$\{{\hat\beta}_{tT}\}$ defined by (\ref{eq:6.3}). It follows that
the enhanced information process $\{{\hat\xi}_t\}$ defined by
(\ref{eq:6.2}) is independent of $\{\delta_t\}$. Furthermore, the
filtration generated jointly by $\{\xi_t,\xi'_t\}$ is equivalent to
that generated jointly by $\{{\hat\xi}_t,\delta_t\}$. Since
$\{\delta_t\}$ provides no useful information about $X_T$, i.e.
$J(\delta_t,X_T)=0$, the informed trader in effect has
$\{{\hat\xi}_t\}$ as the primary basis for valuation.

This line of argument, making use of the orthogonalisation
procedure, extends more generally to the case where there are
multiple information processes relating to the cash flow $X_T$:
Starting with a family of processes $\{\xi_t^k\}_{k=1,\ldots,n}$
with signal-to-noise ratios $\{\sigma_k\}_{k=1,\ldots,n}$ one
orthogonalises the associated noises. The result is a new set of
information processes $\{{\bar\xi}_t^k\}_{k=1,\ldots,n}$ with
signal-to-noise ratios $\{{\bar\sigma}_k\}_{k=1,\ldots,n}$. Then the
information process that the informed trader uses as a basis for
valuation can be represented by a single \textit{effective}
information process $\{{\hat\xi}_t\}$ with the enhanced
signal-to-noise ratio ${\hat\sigma}=(\sum_k
{\bar\sigma}_k^2)^{1/2}$.

\section{Additional information held by the informed trader and
statistical arbitrage strategies exploiting this} \label{sec:7}

We are in a position to quantify the amount of excess information
held by the informed trader above that of the market. This is
measured by the difference of the mutual information:
\begin{eqnarray}
\Delta J = J({\hat\xi}_t,X_T) - J(\xi_t,X_T).
\end{eqnarray}
By the argument in \S\ref{sec:3}, the mutual information of the
informed trader is given by an entropy difference of the form
\begin{eqnarray}
J({\hat\xi}_t,X_T)= H({\hat\xi}_t) - H({\hat\xi}_t|X_T).
\end{eqnarray}
The entropy $H({\hat\xi}_t)$ of the `insider' information is
determined by the marginal density of ${\hat\xi}_t$, whereas the
conditional entropy $H({\hat\xi}_t|X_T)$ is the entropy of a
Brownian bridge.

\begin{figure}
\begin{center}
  \includegraphics[scale=0.8]{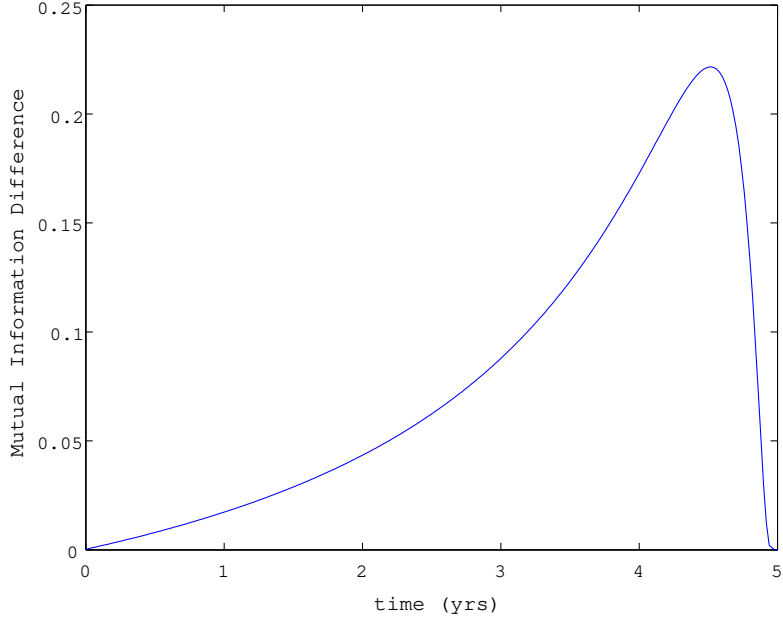}
  \caption{Mutual information difference. The additional information
held by the informed trader over that of the market is nonnegative.
The parameters are set to be $h_1=0$, $h_2=1$, $p_1=0.2$, $p_2=0.8$,
$T=1$, $\sigma=0.25$, $\sigma'=0.45$, and $\rho=0.15$.
  \label{fig:4}
  }
\end{center}
\end{figure}

Following the line of argument presented in \S4 we are able to
represent the mutual information difference in terms of the expected
entropy differences:
\begin{eqnarray}
\Delta J = {\mathbb E}[H_t] - {\mathbb E}[{\tilde H}_t],
\end{eqnarray}
where ${\tilde H}_t = -\sum_i {\tilde\pi}_{it}\ln{\tilde\pi}_{it}$.
This expression makes it apparent that $\Delta J$ is nonnegative,
since the entropy characterises the amount of uncertainty concerning
the value of the cash flow $X_T$, and for any $t\in(0,T)$ this
uncertainty is greater on average for the general market
participant than for the informed trader. In Figure~\ref{fig:4} we
plot an example of $\Delta J$, indicating the way in which the
excess information held by the informed trader changes in time.

Given that the informed trader is on average `more knowledgable'
than the general market participant it is natural to ask how this
advantage can be turned into profit. One of the issues that can be
addressed in this connection is the derivation of optimal trading
strategies. For such an analysis one may need to introduce
additional structure into the problem in the form of a suitable
criterion for optimality and a specification of the market price of
risk. In the present investigation, we confine the discussion to a
demonstration, supported by simulation studies, of how even very
simple strategies can yield statistical arbitrage opportunities by
outperforming the market.

For example, suppose we consider a strategy such that at some
designated time $t\in[0,T]$ a market trader purchases a digital bond iff
at that time the bond price $B_{tT}$ is greater than $KP_{tT}$ for some
specified threshold $K$. The value of $K$ can be regarded as
the risk aversion level of the trader. An informed trader follows
the same rule, but makes a better estimate for the value of the bond, and
hence purchases the bond iff ${\tilde B}_{tT}>KP_{tT}$.
In either case a bond that is purchased will be held until maturity.
That such a strategy leads to a statistical arbitrage opportunity for the
informed trader
can be seen as follows. We assume that the initial position of the
trader is zero, i.e. purchase of a digital bond at $t$ requires
borrowing the amount $B_{tT}$ at that time, and repaying the amount
$P_{tT}^{-1}B_{tT}$ at $T$. Thus the value of the market trader's
portfolio at $T$ is
\begin{eqnarray}
V_T={\mathds 1}{\{B_{tT}>KP_{tT} \}} (X_T-P_{tT}^{-1}B_{tT}),
\end{eqnarray}
whereas the terminal value of the informed trader's portfolio is
\begin{eqnarray}
{\tilde V}_T={\mathds 1}{\{{\tilde B}_{tT}>KP_{tT}\}}
(X_T-P_{tT}^{-1}B_{tT}).
\end{eqnarray}
Consider now the present value $P_{0T} {\mathbb E}[\Delta V_T]$ of a
security that delivers a cash flow equal to the excess P\&L $\Delta
V_T={\tilde V}_T- V_T$ generated by the strategy of the informed
trader. By use of the tower property we have ${\mathbb E}[\Delta
V_T]={\mathbb E}[{\mathbb E}[\Delta V_T|{\mathcal G}_t]]$; but
\begin{eqnarray}
{\mathbb E}[\Delta V_T|{\mathcal G}_t] = P_{tT}^{-1}\left({\mathds
1}{\{ {\tilde B}_{tT}>KP_{tT}\}} - {\mathds 1}{\{B_{tT}>KP_{tT}\}}
\right) \left({\tilde B}_{tT}-B_{tT} \right), \label{eq:7.4}
\end{eqnarray}
since the random variables $B_{tT}$ and ${\tilde B}_{tT}$ are both
${\mathcal G}_t$-measurable. If ${\tilde B}_{tT}>B_{tT}$ then
${\mathds 1}{\{{\tilde B}_{tT}>KP_{tT}\}} - {\mathds
1}{\{B_{tT}>KP_{tT}\}}$ is nonnegative, whereas if ${\tilde
B}_{tT}<B_{tT}$ then ${\mathds 1}{\{{\tilde B}_{tT}>KP_{tT}\}} -
{\mathds 1}{\{B_{tT}>KP_{tT}\}}$ is nonpositive. It follows that
${\mathbb E}[\Delta V_T|{\mathcal G}_t]$ is a nonnegative random
variable, and hence ${\mathbb E}[\Delta V_T]>0$, since ${\mathbb
E}[\Delta V_T|{\mathcal G}_t]>0$ with probability greater than zero.
Therefore, the informed trader can execute a transaction at zero
cost that has positive value, and this is what we mean by
`statistical arbitrage'.

\begin{figure}
\begin{center}
  \includegraphics[scale=0.8]{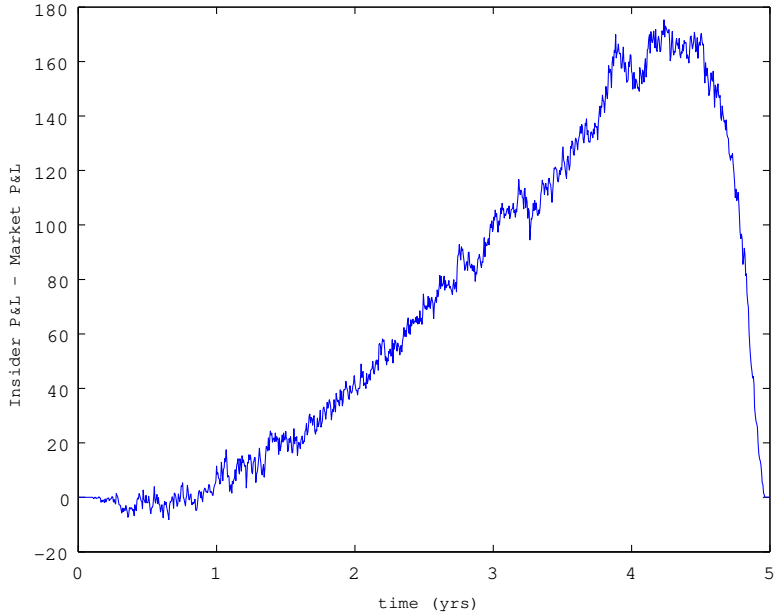}
  \caption{The P\&L difference for digital bonds. At each time the
traders purchase the bond if and only if the valuation of the bond is
greater than a specified threshold. The general market trader buys if
$B_{tT}>KP_{tT}$,
whereas the informed trader uses the condition ${\tilde
B}_{tT}>KP_{tT}$. The difference in profit and loss between the
informed trader and the general market trader is plotted, based on
$2000$ realisations, when the \textit{a priori} probability of
default is $p_1=0.2$. Other parameters are set to be $h_1=0$,
$h_2=1$, $T=5$, $\sigma=0.25$, $\sigma'=0.45$, $\rho=0.15$, and
$K=0.7$.
  \label{fig:5}
  }
\end{center}
\end{figure}

We have examined the profit and loss (P\&L) profile, both for a
general market trader and for an informed trader, resulting from the
repeated application of such a strategy. The results are plotted in
Figure~\ref{fig:5}. In particular, we consider 2000 realisations
(sample paths) for the information processes governing the bond
price valuations. For each
fixed $t\in[0,T]$ we calculate the total P\&L for the informed
trader and for the market trader obtained by following the specified
strategy over and over for each of the 2000 independent sample
paths. For each fixed $t$, we chart in Figure~\ref{fig:5} the
difference between the total P\&L of the informed trader and that of
the market trader. Providing that the strategy is executed after
enough time has passed for the informed trader to gain an
informational advantage, we find that the difference between the
P\&L of the informed trader and that of the market trader is always
positive. Furthermore, the qualitative behaviour of the resulting
P\&L difference is in agreement with the qualitative behaviour of
the magnitude of the excess information possessed by the informed
trader indicated in Figure~\ref{fig:4} (we have chosen the same
parameter values for these two figures to allow for direct
comparison).

Our objective has been to demonstrate how statistical arbitrage
strategies arise in a market characterised by
heterogeneous information flow. It is interesting that a
qualitatively similar behaviour for the excess information and the
excess P\&L is observed in the case of the rather primitive strategy
we have considered here. There are many ways in which one can
improve upon the trading strategy examined above. An important
open issue is to determine the optimal trading strategy, subject to
suitable optimality criteria, that exploits the excess information.
We conclude by remarking that a related approach to the modelling of
insider trading within the information-based framework is suggested
in Macrina (2006), where the asset return is modelled as being
dependent on more than one market factor, for which only some of the
associated information processes are accessible to the market. It
would be of interest to examine whether the kind of hedge fund
strategy considered here is also applicable in a setup involving
multiple market factors.

\begin{acknowledgements}
We thank J.~Z.~Kelly, A.~Macrina, B.~K.~Meister, and M.~F.~Parry for
stimulating discussions.
\end{acknowledgements}

\end{document}